\documentclass[10pt,conference]{IEEEtran} 
 
\usepackage{graphicx}
\usepackage{listings}
\usepackage{amsmath}
\usepackage{amssymb}
\usepackage{float}
\usepackage{array}
\usepackage{hyperref}
\input{Qcircuit.tex}

\begin{document}

\title{ The limits of quantum circuit simulation with low precision arithmetic }
\author{  {Santiago I. Betelu}\\
           University of North Texas, Mathematics Department\\
           Datavortex Technologies\\
           Denton, TX\\
           santiago.betelu@unt.edu\\
       }

\maketitle

\begin{abstract}
This is an investigation of the limits of quantum circuit simulation with Schrodinger's formulation and low precision arithmetic. The goal is to estimate how much memory can be saved in simulations that involve random, maximally entangled quantum states. An arithmetic polar representation of $B$ bits is defined for each quantum amplitude and a normalization procedure is developed to minimize rounding errors. Then a model is developed to quantify the cumulative errors on a circuit of $Q$ qubits and $G$ gates. Depending on which regime the circuit operates, the model yields explicit expressions for the maximum number of effective gates that can be simulated before rounding errors dominate the computation. The results are illustrated with random circuits and the quantum Fourier transform.    
\end{abstract}


\section{Introduction}

Simulating quantum circuits with Schrodinger's formulation is costly because all the
coefficients of the wave function must be permanently stored in memory, thus every qubit added to the computation doubles memory usage, communication bandwidth\footnote{In a cluster system most of the coefficients must be communicated between nodes when computing non diagonal gates.} and computing time. The problem is unavoidable when the simulation handles random maximally entangled states and large depths. 
With current technology, the maximum number of qubits that can be simulated in this situation is around $Q \approx 50$, the quantum supremacy limit\footnote{Alternative simulation methods, such as the path-integral formulation and tensor contraction networks may work well above 50 qubits \cite{IBM2019,Markov} but they are feasible only if the connectivity of the gates is reduced to a grid, or the circuits have small depth, or the simulations handles states close to linear combinations of product states. For the general random states we consider here these methods do not offer an advantage. Their memory and time complexity scalings are summarized in \cite{Aaronson}}.

The goal of this work is to investigate how much memory and communication bandwidth
can be saved in Schrodinger's formulation by using a low precision arithmetic format, and to obtain analytic expressions to demarcate the limits 
of this approach. Throughout this paper it is assumed that the simulated circuit may have unrestricted gate connectivity, 
large depth and generates maximally entangled random states at some point of the evolution.

Consider the simulation of a quantum circuit with $Q$ qubits. At each time step, the quantum state $|\psi\rangle$ is determined by $N=2^Q$  complex coefficients $c_k$, 
\begin{equation}
|\psi\rangle= \sum_{k=0}^{N-1} c_k|k\rangle, 
\label{psi}
\end{equation}
where $|k\rangle$ are the computational basis states and the coefficients are normalized as 
\begin{equation}
\sum_{k=0}^{N-1}|c_k|^2=1.  \label{normalization}
\end{equation}
Each coefficient $c_k$ is stored in $B$ bits,
thus we need $2^QB$ bits in total. Since it is assumed that the states are random and maximally entangled, the real and imaginary parts of the coefficients are random and uniformly distributed on the 
surface of a $2N$-dimensional unit sphere \cite{Boixo}, thus the coefficients are incompressible in the information theoretical sense. 

The strategy is to develop a model that will serve as a guide on how setup these coefficients 
using a low precision format in the most economical way possible as to minimize $B$. 
This depends on the desired accuracy and the number of gates susceptible to truncation error.
In section II the low precision format for $c_k$ is defined in terms of discretized logarithms. 
In section III an analytic
expression is developed for the cumulative error after executing $G$ gates, and this is used to determine the optimal parameters that minimize the error. There it is shown that if the state is sufficiently random, $16$ bits per coefficient are sufficient to obtain useful results for several hundred gates. In Section IV it is shown that when the errors are biased then the total variance may accumulate quadratically, and a normalization procedure is developed to minimize the problem. 
Finally Section V tests the low precision error model with a benchmark of random circuits that generate high entropy entangled states, and with non-random runs of the Quantum Fourier Transform.

%%%i%%%%%%%%%%%%%%%%%%%%%%%%%%%%%%%%%%%%%%%%%%%%%%
\section{Low precision format for the amplitudes}
Since a useful simulation must handle amplitudes that vary by many orders of magnitude, 
it is convenient to define a format based on their complex logarithm $\log c_k$. 
Then we approximate each amplitude $c_k$ with a function $T(c_k)$ and a triplet of integers $(e_k, f_k, a_k)$, of size $E, F$ 
and $A$ bits respectively, with $0\le e_k < 2^E$, $0\le f_k < 2^F$, $0\le a_k < 2^A$,  
\begin{equation}
   c_k \approx T( c_k ) = \exp\left( -\left(e_k + \frac{f_k}{2^F} \right)+ 2\pi i \frac{a_k}{2^{A}} \right),
  \label{format}
\end{equation}
where $i$ is the imaginary unit. The total number of bits per amplitude is $B=E+F+A$, one goal of this work is to find the optimal values for this triplet.
\begin{figure}[ht]
    \centering
    \includegraphics[width=4cm]{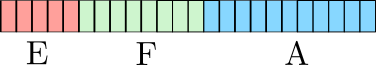}
    \caption{ The complex amplitudes are encoded with $E$ bits for the integer part of the exponent, $F$ bits for the fraction and $A$ bits for the argument. 
    \label{lowprecisionsketch}}
\end{figure}
This is a storage format and not a native arithmetic format, thus to perform 
all calculations we convert to and from double precision numbers and perform all arithmetic 
operations with double precision. To convert a double precision complex $c_k$ to low precision complex,
\begin{equation}
2^F e_k + f_k=  \left[ -2^F  \ln |c_k|  \right],  \;\;\;\;\; a_k= \left[  \frac{ \arg(c_k)}{2\pi} 2^A \right]
\label{T}
\end{equation} 
where one must take care that $\arg(c_k)$ is defined in the interval $[0,2\pi)$ and the brackets indicate nearest integer rounding.

The smallest nonzero modulus $|c_k|$ encoded by this format occurs when $e_k=2^E-1$ and $f_k=2^F-1$,
\begin{equation}
\mu= \min |c_k|= \exp(-2^E+2^{-F})       \label{mu}  
\end{equation}
and the largest $|c_k|=1$ occurs when $e_k=f_k=0$.
The word $e_k= 2^E-1$, $f_k=2^F-1$ and $a_k=2^A-1$ (all binary digits equal to one) is reserved for the underflow numbers including $c_k=0$, and for all these, 
\begin{equation}
T (c_k)=0 \;\;\;\ \mbox{if } \;\; |c_k|<\mu.   
\end{equation}

This format has some advantages: a) It is more regular than floating point numbers in which the distance 
between consecutive numbers jumps by a factor $2$ each time the exponent 
increases, while in the logarithmic format it changes smoothly with $e_k, f_k$ and $a_k$. b) The most important 
advantage is that for random amplitudes, the rounding errors of $\log c_k$ are uniformly distributed. This feature  
facilitates the analysis of the errors in the sections that follow. c) Phase gates (controlled or otherwise) with angle $\pi/2^k$ can be computed without error 
if $k<A$. d) The format uses a single 
exponent $e_k$ for the whole complex number, saving a few bits (complex floats need two exponents).
The main drawback is that modern CPUs do not have circuitry to handle this format natively, 
but it is possible to optimize code for the conversion to and from floating point numbers by using lookup tables and interpolation.

In order to illustrate the use of the low precision format, we compute the evolution of $|\psi\rangle$ with a simplified version of the 
parallel algorithm developed in \cite{Raedt} 
that employs an ingenious notation to describe the implementation of the gates. 
Table \ref{tablegates} shows how the operations are performed on the vector of amplitudes $c_k=c(k)$, where the index $k$ is written in binary between parentheses, the bit at position $p$ is denoted by $0_p$ or $1_p$, and the dots represent unaffected bits. 
The main difference is that here, the $N$ coefficients $c_k$ are stored in memory with low precision (the words $e_k,f_k,a_k$ of $T(c_k)$), and these are converted to double precision complex to perform all the arithmetic operations according to Table \ref{tablegates}, and then the results are converted back to low precision. The reader can find the C/MPI implementation used in this work in \cite{mygithub}.
\begin{table}[h]
	\renewcommand{\arraystretch}{1.5}
	\centering
	\begin{tabular}{|l|l|}
		\hline
		{\bf Gate }                          &  {\bf Operation} \\
		\hline
		$H(q)$ &  $c(.., 0_q,..)\leftarrow \frac{1}{\sqrt{2}}\left(c(.., 0_q, ..)+c(.., 1_q,..)\right)$ \\
		&  $c(.., 1_q,...)\leftarrow \frac{1}{\sqrt{2}}\left(c(.., 0_q, ..)-c(.., 1_q,..)\right)$ \\
	%	\hline
	%	$U_3(\theta,\lambda,\phi,q)$ &  $c(.., 0_q,..)\leftarrow c(.., 0_q, ..)\cos\frac{\theta}{2}-c(.., 1_q,..)e^{i\lambda}\sin\frac{\theta}{2}$ \\
	%	                             & $c(.., 1_q,...)\leftarrow  c(.., 0_q, ..)e^{i\phi}\sin\frac{\theta}{2}-c(.., 1_q,..)e^{i(\phi+\lambda)}\cos\frac{\theta}{2} \\
		\hline
		CNOT $(p,q)$        &  $c(.., 1_p,.., 0_q, ..)\leftrightarrow c(.., 1_p,.., 1_q, ..)$ \\
		\hline
		CP$(p,q)$ &  $c(.., 1_p, .., 1_q, ..)\leftarrow e^{i \pi/2^m} c(.., 1_p, .., 1_q, ..)$ \\
		\hline
		SWAP$(p,q)$         &  $c(.., 1_p,.., 0_q, ..)\leftrightarrow c(.., 0_p,.., 1_q, ..)$ \\
		\hline
	\end{tabular}
	\vspace{1mm} 
	\caption{How gates operating on qubits $p,q$ are implemented:  "$\leftarrow$" represents assignment and "$\leftrightarrow$" swapping. Parentheses contain the binary index $k$ of $c_k$ and the dots indicate unaffected bits. $H$ is Hadamard's gate and CP are the controlled phase gates.        \label{tablegates}}
\end{table}

%%%%%%%%%%%%%%%%%%%%%%%%%%%%%%%%%%%%%%%%%%%%%%%%%%%%%%%%%%%%%%%%%%%%%
\section{Total conversion error for unbiased uniformly distributed independent errors}
Now we compute the total conversion error for maximally entangled random states and later we will use this result to find the optimal values of $E,F$ and $A$.

\begin{figure}[ht]
	\centering
	\includegraphics[width=8.8cm]{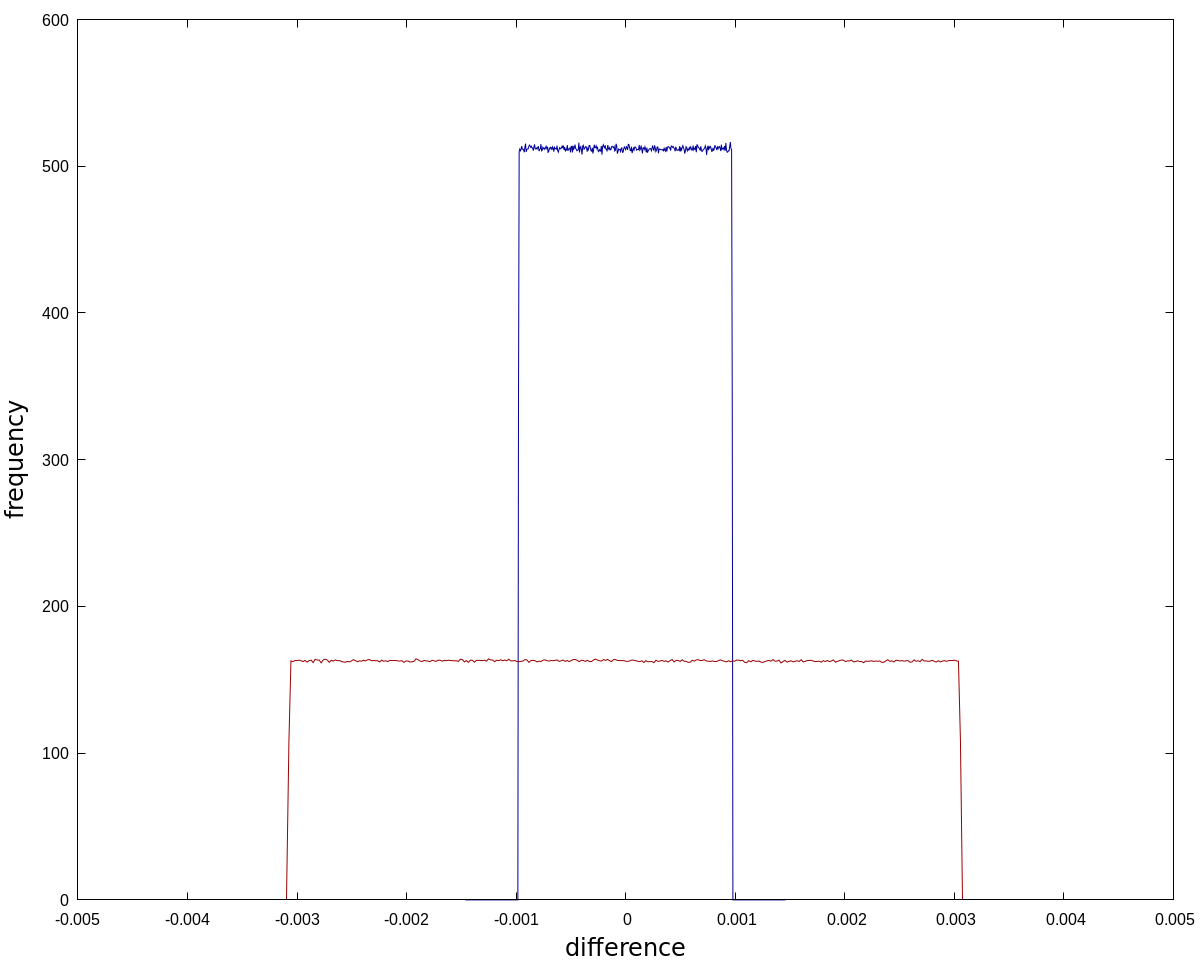}
	\label{histograms}
	\caption{Empirical histograms of the rounding errors for the logarithm of the modulus (high rectangle) and for the argument of Eq. (\ref{format}) for $Q=20$, $E=5, F=9$ and $A=10$. They are uniformly 
		distributed when the real and imaginary parts of the coefficients $c_k$ are 
		random because the rounded binary digits after the least significant digit are random. This is not true for floating point formats. 
		\label{histogramrounding}}
\end{figure}

Suppose we have a double precision vector state $|\psi \rangle$ and we convert all the coefficients using format Eq. (\ref{format}) and denote the 
result as $T(|\psi\rangle)$. 
Let $z_k= \log c_k$ be the complex logarithm of the exact amplitude and $\epsilon_k+i\gamma_k=\log T(c_k)-\log c_k$ the error of the complex logarithm when we convert to low precision and round all the binary digits beyond the last significant bits of $z_k$ (the bits at positions $F$ for the register $f_k$ and $A$ for the register $a_k$).  
An advantage of the low precision format is that, if the coefficients $c_k$ are random, the distribution of the errors of the logarithm is uniform and bounded 
on the intervals $-2^{-F}/2 \le \epsilon_k \le 2^{-F}/2$ and $-\pi 2^{-A}\le \gamma_k\le \pi 2^{-A}$. This is illustrated in
Fig. (\ref{histogramrounding}), it depicts the empirical histograms of the errors $\epsilon_k$ and $\gamma_k$ for normally distributed amplitudes $c_k$. This feature (which does not occur with floating point numbers) will greatly simplify the computations that follow.

From the definition of the errors, $T(c_k)=e^{z_k+ \epsilon_k+i\gamma_k}$ and the total quadratic error of the $N$ terms adds up to
\[
\varepsilon_c^2= \| T |\psi\rangle - |\psi\rangle   \|^2 = \sum_{k=0}^{N-1} |T(c_k)- c_k|^2 =
 \]
 \[
  \sum_{|c_k|< \mu} |0-c_k|^2+ \sum_{|c_k|\ge\mu} \left| e^{z_k+ \epsilon_k+i\gamma_k} - e^{z_k} \right|^2 
\]
\begin{equation}
\varepsilon_c^2= \sum_{|c_k|< \mu} |c_k|^2+ \sum_{|c_k|\ge \mu } |c_k|^2 \left| e^{\epsilon_k+i\gamma_k} - 1 \right|^2
\label{roundingerror}
\end{equation}
where we have split the terms that underflow ($|c_k|<\mu$ and  $T(c_k)=0$, see Eq. (\ref{mu})) and those that don't ($|c_k|\ge\mu$). 

To compute the expected value of this error we assume a random state in which the $c_k$ are uniformly distributed on the surface of a unit sphere of $2N$ dimensions, 
and then the probabilities $p=|c_k|^2$ are distributed according to Porter-Thomas distribution with PDF $f(p)\approx Ne^{-pN}$ \cite{Boixo}. 
Now we compute the expected value of the first sum (last line) of Eq. (\ref{roundingerror}), 
\[
\phi= \mathbb{E}\left( \sum_{|c_k|< \mu} |c_k|^2 \right)\approx N\int_0^{\mu^2} pNe^{-Np} dp 
\]
\begin{equation}
\phi \approx 1-(N\mu^2+1)e^{-N\mu^2}.
\label{phi}
\end{equation}
This number represents how much the normalization condition Eq. (\ref{normalization}) drops because of the underflows. Just to give an idea of the scale, $\phi$ is approximately $N^2 \mu^4/2$ when $N\mu^2\ll 1$, thus if $E$ is sufficiently large this number is very small. 

Now the second sum of Eq. (\ref{roundingerror}),
if the errors are uniformly distributed and independent of the coefficients themselves, then one can compute the expected value of the factor inside the sum,
\begin{equation}
    {\mathbb E} \left| e^{ \epsilon_k +i\gamma_k } -1 \right|^2= 
    \frac{1}{\epsilon\gamma}\int_{-\epsilon/2}^{\epsilon/2} \int_{-\gamma/2}^{\gamma/2} \left| e^{ \epsilon_k +i\gamma_k } -1 \right|^2 \; d\epsilon_k \; d\gamma_k 
\end{equation}
\[\approx \frac{ 2^{-2F} +  4\pi^2 2^{-2A}}{12},\]
where $\epsilon= 2^{-F}$ and $\gamma=2\pi 2^{-A}$ and we only kept leading order terms. 

Substituting into Eq. (\ref{roundingerror}), using $\sum_{|c_k|\ge \mu } |c_k|^2=1-\phi$ and taking the constant expected value out of the sum, we obtain the expected value of the conversion error for random states
\begin{equation}
\varepsilon_c^2 \approx \phi + (1-\phi)\frac{ 2^{-2F} +  4\pi^2 2^{-2A}}{12}.
\label{conversionerror}
\end{equation}
The value of $\phi$ (Eq. (\ref{phi})) is always much smaller than unity when the values of $E,F,A$ are optimal. In the following section we compute the optimal values of the parameters.

\subsection{Optimal triplets $E,F,A$}

The computation of the optimal values of the parameters $E,F,A$ can be performed by solving the discrete minimization of Eq. (\ref{conversionerror}) with Eqs. (\ref{phi}),(\ref{mu}), which is straightforward to do by testing all combinations of $E,F,A$ (Table \ref{EFArandom}).
\begin{table}[h]
	\centering
	\begin{tabular}{|l| l | l| l| l| }
		\hline
		& $Q=20$ &  $Q=30$ & $Q=40$ & $Q=50$  \\
		\hline
		$B$ &  $E, F, A$ &   $E, F, A$ &  $E, F, A$ & $E,F,A$  \\
		\hline
		8  & 4, 1, 3  & 4, 1, 3  & 4, 1, 3  & 5, 0, 3  \\ 
		9  & 4, 1, 4  & 4, 1, 4  & 4, 1, 4  & 5, 1, 3  \\ 
		10  & 4, 2, 4  & 4, 2, 4  & 4, 2, 4  & 5, 1, 4  \\ 
		11  & 4, 2, 5  & 4, 2, 5  & 4, 2, 5  & 5, 2, 4  \\ 
		12  & 4, 3, 5  & 4, 3, 5  & 4, 3, 5  & 5, 2, 5  \\ 
		13  & 4, 3, 6  & 4, 3, 6  & 4, 3, 6  & 5, 3, 5  \\ 
		14  & 4, 4, 6  & 4, 4, 6  & 4, 4, 6  & 5, 3, 6  \\ 
		15  & 4, 4, 7  & 4, 4, 7  & 4, 4, 7  & 5, 4, 6  \\ 
		16  & 4, 5, 7  & 4, 5, 7  & 4, 5, 7  & 5, 4, 7  \\ 
		17  & 4, 5, 8  & 4, 5, 8  & 4, 5, 8  & 5, 5, 7  \\ 
		18  & 4, 6, 8  & 4, 6, 8  & 5, 5, 8  & 5, 5, 8  \\ 
		19  & 4, 6, 9  & 4, 6, 9  & 5, 6, 8  & 5, 6, 8  \\ 
		20  & 4, 7, 9  & 4, 7, 9  & 5, 6, 9  & 5, 6, 9  \\ 
		21  & 4, 7, 10  & 4, 7, 10  & 5, 7, 9  & 5, 7, 9  \\ 
		22  & 4, 8, 10  & 4, 8, 10  & 5, 7, 10  & 5, 7, 10  \\ 
		23  & 4, 8, 11  & 4, 8, 11  & 5, 8, 10  & 5, 8, 10  \\ 
		24  & 4, 9, 11  & 4, 9, 11  & 5, 8, 11  & 5, 8, 11  \\ 
		25  & 4, 9, 12  & 4, 9, 12  & 5, 9, 11  & 5, 9, 11  \\ 
		26  & 4, 10, 12  & 4, 10, 12  & 5, 9, 12  & 5, 9, 12  \\ 
		27  & 4, 10, 13  & 4, 10, 13  & 5, 10, 12  & 5, 10, 12  \\ 
		28  & 4, 11, 13  & 4, 11, 13  & 5, 10, 13  & 5, 10, 13  \\ 
		29  & 4, 11, 14  & 4, 11, 14  & 5, 11, 13  & 5, 11, 13  \\ 
		30  & 4, 12, 14  & 4, 12, 14  & 5, 11, 14  & 5, 11, 14  \\ 
		31  & 4, 12, 15  & 4, 12, 15  & 5, 12, 14  & 5, 12, 14  \\ 
		32  & 4, 13, 15  & 4, 13, 15  & 5, 12, 15  & 5, 12, 15  \\ 
		33  & 4, 13, 16  & 4, 13, 16  & 5, 13, 15  & 5, 13, 15  \\ 
		34  & 4, 14, 16  & 4, 14, 16  & 5, 13, 16  & 5, 13, 16  \\ 
		35  & 4, 14, 17  & 4, 14, 17  & 5, 14, 16  & 5, 14, 16  \\ 
		36  & 4, 15, 17  & 4, 15, 17  & 5, 14, 17  & 5, 14, 17  \\ 
		37  & 4, 15, 18  & 4, 15, 18  & 5, 15, 17  & 5, 15, 17  \\ 
		38  & 4, 16, 18  & 5, 15, 18  & 5, 15, 18  & 5, 15, 18  \\ 
		39  & 4, 16, 19  & 5, 16, 18  & 5, 16, 18  & 5, 16, 18  \\ 
		40  & 4, 17, 19  & 5, 16, 19  & 5, 16, 19  & 5, 16, 19 \\	
		\hline
	\end{tabular}
    
	\vspace{1mm}
	\caption{Optimal triplets  $E,F,A$ with respect of the expected value of the conversion error for random states, computed by brute force minimization of Eq. (\ref{conversionerror}) using (\ref{phi}),(\ref{mu}) with the constraint $E+F+A=B$. \label{EFArandom}}
\end{table}

By minimizing the error with respect to $F,A$ with the constraint $E+F+A=B$
and using Lagrange multipliers, we obtain the optimal relationship between $F$ and $A$
\begin{equation}
A\approx  F + \log_2 (2\pi) ,  \label{opta}
\end{equation}
which must be rounded to an integer, depending on whether the total number of bits $E+F+A$ is even or odd,
$A=F+2$ or $A=F+3$. 

\subsection{Error accumulation after multiple gates with uniform, identically distributed unbiased errors}
During the simulation of a quantum circuit, the main source of error occurs during the conversion from the double precision format where the actual arithmetic is performed, and this error accumulates after the computation of multiple error-prone gates.  
Next we derive an approximation of the cumulative error after $G$ error-prone conversions 
\begin{equation}
\sigma^2=\|\; |\psi_G\rangle - |\psi_{ex}\rangle \; \|^2,
\label{cumulative}
\end{equation}
where $|\psi_{ex}\rangle$ denotes the exact state and $|\psi_{G}\rangle$ the low precision one. To give a sense of the scale of $\sigma^2$, it is related to the pure-state fidelity
$\Phi=|\langle \psi_G | \psi_{ex} \rangle|^2$ as $\Phi\ge \left( 1-\sigma^2/2 \right)^2.$ 
Thus, for example, $\sigma^2=1/4$ represents a fidelity of $\Phi \ge 0.765$.

Let $|\varepsilon_t\rangle=|\psi_G(t)\rangle - |\psi_{ex}(t)\rangle$ be the cumulative error vector at time step $t$, and $|\tau_t\rangle$ the conversion error, which variance is given by Eq. (\ref{conversionerror}). At the next time step we apply an arbitrary unitary gate $U_t$ to the state $|\psi_G\rangle$ and then perform a conversion $T$ 
\begin{equation}
    |\psi_G(t+1)\rangle = U_t (|\psi_{ex}(t)\rangle +|\varepsilon_t\rangle ) +|\tau_t\rangle 
\end{equation}
thus the new error is 
\begin{equation}
|\varepsilon_{t+1}\rangle= U_t|\varepsilon_t\rangle +|\tau_t\rangle.
\end{equation}
Because the gate is unitary, $ \| U_t|\varepsilon_t\rangle \|^2 = \| |\varepsilon_t\rangle \|^2 $, and
assuming the errors are random, independent, identically distributed and unbiased, the variances can be added for the new error
$\| |\varepsilon_{t+1}\rangle \|^2 \approx \| |\varepsilon_t\rangle \|^2+ \| |\tau_{t}\rangle \|^2$, thus the cumulative variance increases linearly as a first approximation.
That is, we can use the central limit theorem to find the cumulative error after $G$ gates as the sum of the variances of $G$ random variables each with variance given by Eq. (\ref{conversionerror}), 
\begin{equation} 
\sigma^2 \approx \left(\phi+(1-\phi)\frac{2^{-2F}+4\pi^2 2^{-2A}}{12} \right) G. 
\label{iid}
\end{equation}

Then one can estimate the maximum number of gates that can be computed before the error reaches the tolerance $\sigma^2$ with Eq. (\ref{iid}),
\begin{equation}
G_{random}< \frac{\sigma^2}{ \varepsilon_c^2 },
\label{iidgates}
\end{equation}
where $\varepsilon_c^2$ is the conversion error Eq. (\ref{conversionerror}).
The first and third columns of Table \ref{Gtable} shows how many gates can be simulated as a function of the word size $B$ for a barely tolerable simulation with $\sigma=1/2$ (fidelity $\ge0.765$). 

\subsection{Effective number of error-prone gates}
Not all gates are susceptible to rounding error, for example NOT, CNOT, SWAP and phase gates with angle $\pi/2^{k}$, $k<A$ are error-free because they basically involve memory 
swaps, sign changes or integer changes of the register $a_k$. As for the error-prone gates ($H, X^{1/k}$, etc.) their degree of error contribution depends on details such as the number of controls. Here we  approximate the effective number of error-prone gates by the cumulative fraction of coefficients directly modified by the gates, 
\begin{equation}
G=  \sum_{g=1}^{G_0} \beta_g \label{Gp},
\end{equation}
where $G_0$ is the total number of gates, $\beta_g$ is the fraction of coefficients affected by gate $g$, given in Table \ref{effectiveG}.
\begin{table}[h]
    \renewcommand{\arraystretch}{1.5}
    \centering
    \begin{tabular}{| p{6.5cm} | p{0.7cm}|  }
        \hline
        {\bf Gate type }                           &  {\bf $\beta_g$} \\
        \hline
        $X, Z^{1/k} \ (k<A)$, CNOT, SWAP, TOFF  &  $0$  \\
        \hline    
         $Z^{1/k} \ (k\ge A)$     &  $1/2$   \\
        \hline
        H, $X^{1/k}$, $Y^{1/k} \; (k> 2) , U_3(\theta,\lambda,\phi)$           &  $1$ \\
        \hline
        Last row with $k$ controls                  &  $1/2^k$ \\ 
        \hline
    \end{tabular}
    
    \vspace{1mm}
    \caption{Fraction of coefficients affected by rounding error for typical gates. \label{effectiveG}}
\end{table}
The logic behind these fractions is that in Eq. (\ref{roundingerror}) only the terms with coefficients directly modified by the gates can be non-zero.
This is only an approximation because for example, phase gates only introduce error on the complex argument (the angle), but this is sufficient to obtain a rough estimate of G.
Depending on the algorithm being simulated, the number of noisy gates may be much smaller than the actual number of gates.  

\section{Biased errors}

\subsection{Upper bound for the total conversion error}
Unlike the errors studied in previous sections, biased errors do not tend to cancel each other. 
Starting again with Eq. (\ref{roundingerror}) and using the normalization condition Eq. (\ref{normalization}) it is possible to obtain a simple expression for the upper
bound of the conversion error,
\begin{equation}
\varepsilon_b^2=\| T |\psi\rangle - |\psi\rangle \|^2   \le N\mu^2 +  \left| e^{2^{-F-1}+ 2\pi i 2^{-A-1} } -1 \right|^2,      
\end{equation}
where we have taken the values $\epsilon_k= 2^{-F-1}$, $\gamma_k= \pi 2^{-A}$ that maximize each term of 
the error (when $A,F$ are larger than $2$), and included the underflow term
$N \mu^2$ by considering the worst case scenario where all $N$ terms in Eq. (\ref{roundingerror}) contribute to the maximum underflow value $|c_k|^2=\mu^2$ (some terms are counted twice but that is OK because we are computing an upper bound). Then, by dropping higher order terms,
\begin{equation}
\varepsilon_b^2=\| T |\psi\rangle - |\psi\rangle   \|^2   \lesssim N\mu^2 + \frac{2^{-2F}+4\pi^2 2^{-2A}}{4} . 
\label{conversionbound}
\end{equation}

After executing $G$ error-prone gates, we  consider the worst case scenario where all the errors are constant, then the total error is just a multiple of $\varepsilon_b$,
\begin{equation}
\sigma   \lesssim G\sqrt{ N\mu^2 + \frac{2^{-2F}+4\pi^2 2^{-2A}}{4} }  . 
\label{noassumptions}
\end{equation}
 That means that now, $\sigma^2$ grows as $O(G^2)$. Generally, this error is much larger than than Eq. (\ref{iid}) and then one can execute far fewer gates than with the random case (compare with Eq.(\ref{iidgates})),
\begin{equation}
G_{biased}< \frac{\sigma}{ \sqrt{\varepsilon_b^2} }.
\label{biasedgates} 
\end{equation}
In next section it is shown how to partially remediate this biased situation by multiplying the coefficients by carefully crafted factors that bring the errors closer to Eq.(\ref{iid}) than to Eq.(\ref{noassumptions}).
 
It is also possible to compute optimal values of the parameters $E,F,A$ that minimize the upper bound of the error Eq. (\ref{conversionbound}) (Table \ref{EFAbiased}).
For these optimal triplets, it is also possible to compute the conversion error for random and biased states for $Q=50$ (Table \ref{Gtable}) where it is also predicted how many error-prone gates (Eqs. (\ref{iidgates},\ref{noassumptions})) would be possible be run with 50 qubits and $\sigma=1/2$ (i.e. a barely tolerable computation with fidelity $\Phi\ge 0.765$). The table makes evident that the random case is much more favorable than the biased one.
 
\begin{table}[h]
	\centering
	\begin{tabular}{|l| l | l| l| l|  }
		\hline
		& $Q=20$ &  $Q=30$ & $Q=40$ & $Q=50$  \\
\hline
$B$ &  $E, F, A$ &   $E, F, A$ &  $E, F, A$ & $E,F,A$  \\
		\hline
		8  & 4, 1, 3  & 4, 1, 3  & 4, 1, 3  & 5, 0, 3  \\ 
		9  & 4, 1, 4  & 4, 1, 4  & 4, 1, 4  & 5, 1, 3  \\ 
		10  & 4, 2, 4  & 4, 2, 4  & 4, 2, 4  & 5, 1, 4  \\ 
		11  & 4, 2, 5  & 4, 2, 5  & 4, 2, 5  & 5, 2, 4  \\ 
		12  & 4, 3, 5  & 4, 3, 5  & 5, 2, 5  & 5, 2, 5  \\ 
		13  & 4, 3, 6  & 4, 3, 6  & 5, 3, 5  & 5, 3, 5  \\ 
		14  & 4, 4, 6  & 4, 4, 6  & 5, 3, 6  & 5, 3, 6  \\ 
		15  & 4, 4, 7  & 4, 4, 7  & 5, 4, 6  & 5, 4, 6  \\ 
		16  & 4, 5, 7  & 4, 5, 7  & 5, 4, 7  & 5, 4, 7  \\ 
		17  & 4, 5, 8  & 4, 5, 8  & 5, 5, 7  & 5, 5, 7  \\ 
		18  & 4, 6, 8  & 4, 6, 8  & 5, 5, 8  & 5, 5, 8  \\ 
		19  & 4, 6, 9  & 4, 6, 9  & 5, 6, 8  & 5, 6, 8  \\ 
		20  & 4, 7, 9  & 4, 7, 9  & 5, 6, 9  & 5, 6, 9  \\ 
		21  & 4, 7, 10  & 4, 7, 10  & 5, 7, 9  & 5, 7, 9  \\ 
		22  & 4, 8, 10  & 5, 7, 10  & 5, 7, 10  & 5, 7, 10  \\ 
		23  & 4, 8, 11  & 5, 8, 10  & 5, 8, 10  & 5, 8, 10  \\ 
		24  & 4, 9, 11  & 5, 8, 11  & 5, 8, 11  & 5, 8, 11  \\ 
		25  & 4, 9, 12  & 5, 9, 11  & 5, 9, 11  & 5, 9, 11  \\ 
		26  & 4, 10, 12  & 5, 9, 12  & 5, 9, 12  & 5, 9, 12  \\ 
		27  & 4, 10, 13  & 5, 10, 12  & 5, 10, 12  & 5, 10, 12  \\ 
		28  & 4, 11, 13  & 5, 10, 13  & 5, 10, 13  & 5, 10, 13  \\ 
		29  & 4, 11, 14  & 5, 11, 13  & 5, 11, 13  & 5, 11, 13  \\ 
		30  & 4, 12, 14  & 5, 11, 14  & 5, 11, 14  & 5, 11, 14  \\ 
		31  & 4, 12, 15  & 5, 12, 14  & 5, 12, 14  & 5, 12, 14  \\ 
		32  & 5, 12, 15  & 5, 12, 15  & 5, 12, 15  & 5, 12, 15  \\ 
		33  & 5, 13, 15  & 5, 13, 15  & 5, 13, 15  & 5, 13, 15  \\ 
		34  & 5, 13, 16  & 5, 13, 16  & 5, 13, 16  & 5, 13, 16  \\ 
		35  & 5, 14, 16  & 5, 14, 16  & 5, 14, 16  & 5, 14, 16  \\ 
		36  & 5, 14, 17  & 5, 14, 17  & 5, 14, 17  & 5, 14, 17  \\ 
		37  & 5, 15, 17  & 5, 15, 17  & 5, 15, 17  & 5, 15, 17  \\ 
		38  & 5, 15, 18  & 5, 15, 18  & 5, 15, 18  & 5, 15, 18  \\ 
		39  & 5, 16, 18  & 5, 16, 18  & 5, 16, 18  & 5, 16, 18  \\ 
		40  & 5, 16, 19  & 5, 16, 19  & 5, 16, 19  & 5, 16, 19 \\
		\hline
	\end{tabular}
	\vspace{1mm}
	\caption{Optimal triplets $E,F,A$ for the upper bound of the conversion error  computed by minimizing Eq. (\ref{conversionbound}) with the constraint $E+F+A=B$. Notice how close are these values to the random case Table \ref{EFArandom}.\label{EFAbiased}}
\end{table}

\begin{table}[h]
	\centering
	\begin{tabular}{| l   | l  l   | l l | }
		\hline
		&  Random && Upper bound &\\
		\hline
		$B$ &  $\varepsilon_c^2$ & $G_{random}$ & $\varepsilon_b^2$ & $G_{biased}$ \\
		\hline		
8 &     1.35e-01 & 2  &     4.04e-01 & 1 \\ 
12 &     8.42e-03 & 30  &     2.53e-02 & 3 \\ 
16 &     5.26e-04 & 475  &     1.58e-03 & 13 \\ 
20 &     3.29e-05 & 7600  &     9.87e-05 & 50 \\ 
24 &     2.06e-06 & 121599  &     6.17e-06 & 201 \\ 
28 &     1.28e-07 & 1.94e+06  &     3.85e-07 & 805 \\ 
32 &     8.03e-09 & 3.11e+07  &     2.41e-08 & 3221 \\ 
36 &     5.02e-10 & 4.98e+08  &     1.51e-09 & 12884 \\ 
40 &     3.14e-11 & 7.96e+09  &     9.43e-11 & 51491 \\
    \hline
    \end{tabular}
	\vspace{1mm}
	\caption{Typical values of one-conversion errors $\varepsilon_c^2$ and maximum number of error prone gates for $\sigma=1/2$ and $Q=50$ for random states (Eqs. (\ref{conversionerror},\ref{iidgates})), using the optimal triplets. The last two columns represent the biased worst case scenario (Eqs. (\ref{conversionbound},\ref{biasedgates}).  \label{Gtable}}
\end{table}

\subsection{Normalizing the amplitudes} \label{biased1}
The accumulation of rounding errors can upset the normalization condition Eq. (\ref{normalization}). By using $T(c_k)=e^{z_k+\epsilon_k+i\gamma_k}$ and computing modulus square, instead of unity the norm is found to lie in the interval,
\begin{equation}
\exp\left( - 2^{-F} \right) \le \;\; \|T |\psi\rangle \|^2 = \sum_{k=0}^{N-1} |c_k|^2  e^{2\epsilon_k} \;\; \le \exp\left(  2^{-F} \right).
\end{equation}
The inequality achieves the extreme values when all $\epsilon_k= \pm 2^{-F-1}$ are equal. 
When the errors are systematic (for example when most errors $\epsilon_k$ have the same sign)
the effect will compound as we compute multiple gates, and after around $2^F$ gates the computation may be ruined.

We may still want to simulate circuits where the errors are systematic, in which 
case the obvious solution appears to be dividing the 
amplitudes by the norm $\| |\psi\rangle \|$. This will fail when the norm is so close to unity that the low precision format is unable to distinguish the denominator from $1$. 

\begin{figure}[ht]
    \centering
    \includegraphics[width=8cm]{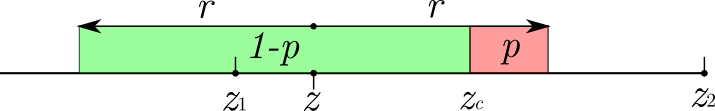}
    \caption{ Let $z=\ln \frac{|c_k|}{\| |\psi\rangle \|}$ and $z_1<z_2$ be two consecutive discrete logarithms with separation $z_2-z_1=2^{-F}=2r$ and $z_1<z<z_2$. We want to round $z$ to the closest of $z_1$ or $z_2$.  After we add a uniformly distributed random number $\delta$ to $z$, with $-r\le \delta <r$, the numbers to the right of $z_c=(z_1+z_2)/2$ are rounded to $z_2$ with probability $p=(z-z_1)/(2r)$ and the numbers to the left of $z_c$ are rounded to $z_1$ with probability $1-p$, thus  ${\mathbb E} ( {\mbox round}( z+\delta)) = (1-p)z_1+pz_2=z$.    \label{randomdiagram}}
\end{figure}

An effective solution is to multiply each coefficient by the exponential of a small independent and uniformly distributed random number, $-2^{-F-1} <\delta<2^{-F-1}$ before converting to low precision,
\begin{equation}
c'_k= \frac{c_k}{  \| |\psi\rangle \| }e^{\delta} \ \ \rightarrow \ \ {\mathbb E} ( T(c'_k))\approx \frac{c_k}{\| |\psi\rangle \| }.
\label{normalize}
\end{equation}
This procedure adds a small $\delta$ to the logarithm of the modulus $z=\ln \frac{|c_k|}{\| |\psi\rangle \|}$.  This works because during conversion, the sum $z+\delta$ is rounded to one of the nearest neighbors $z_1, z_2$ with just the right probability so that the expected value of the logarithm of the conversion modulus equals the desired value, ${\mathbb E}\ln |T(c'_k)|= \ln \frac{|c_k|}{\| |\psi\rangle \|}$ as illustrated in Fig. (\ref{randomdiagram}). 

It is not necessary to perform this operation at each gate, but only if the normalization condition departs from unity by a prescribed fraction. It was found empirically that when the coefficients of $|\psi\rangle$ are random, this procedure is not necessary at all.

\subsection{Systematic rounding errors with small-angle gates} \label{biased2}
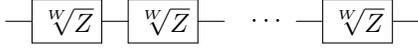
\begin{figure}[ht]
\[ 
   \Qcircuit @C=1em @R=.7em {
 &  &  \gate{\sqrt[W]Z} &  \gate{\sqrt[W]Z} &\qw & \push{\cdots\;\; }  & \gate{\sqrt[W]Z} &\qw \\ 
}
\]
\caption{ A potentially failing circuit after $W>2^A$ applications of the phase gate. The problem is solved by multiplying the amplitudes with appropriate factors. \label{Z}}
\end{figure}

The same argument can be applied to the angular component of the coefficients. This is important for circuits dominated by gates that rotate the qubits a very small angle on the Bloch sphere (like the QFT), where the systematic rounding truncation error can accumulate fast. For the sake of clarity, consider a simple toy circuit composed of $W>2^A$  phase
gates $\sqrt[W]Z$  (phase rotations of $\pi/W$) applied to the same qubit, which should produces a Z gate at the end (Fig. (\ref{Z})).
This circuit will fail because the rotation is not resolved by the low precision format, i.e. the phase change will disappear behind the least significant bit, and the rounding of the conversion operation Eq. (\ref{T}) will leave the coefficients unchanged. 
What happens is that one cannot assume that the errors 
between gates are independent and random, and thus Eq. (\ref{iid}) is not valid anymore and now the 
variance growths as the square of the error-prone-gates as in Eq. (\ref{noassumptions}).
Following the previous subsection, this problem can be solved by rotating the arguments of $c_k$ with small uniform random angles, $-2\pi 2^{-A-1} <\lambda_k<2\pi 2^{-A-1}$ as
\begin{equation}
c'_k= c_k e^{i\lambda_k},
\label{argcorrection}
\end{equation}
where $c_k$ is the double precision complex we want to convert.
Now  the expected value of the argument of the amplitudes is close to the desired value and then ${\mathbb E} ( T(c'_k)) \approx c_k$. This operation only need to be performed right after phase gates with angles less or equal than $\pi/2^A$, or phases that are not of the form $\pi/2^k$.

%%%%%%%%%%%%%%%%%%%%%%%%%%%%%%%%%%%%%%
\section{Testing the model with representative circuits}

The goal of the following tests is to compare the theoretical prediction of the model with the output of two representative simulations: a circuit that generates random states, and a simple quantum Fourier transform with non-random data.

\subsection{Random circuits}
First we test the model for the accumulation of errors with a simple 
circuit that quickly generates random uniformly distributed amplitudes.
\begin{table}[ht]
\centering
\begin{tabular}{|c|}
    \hline                                
    \verb| for i=1,C                               |\\
    \verb|     for q=1,Q                           |\\
    \verb|         k= Q*i+q                        |\\
    \verb|         U3(q, t(k), l(k), p(k) )        |\\
    \verb|         CNOT(q, (q+1)%Q )               |\\
    \verb|     end                                 |\\
    \verb| end                                     |\\  
    \hline
\end{tabular}  
\vspace{2mm}
\caption{Pseudo code of the random circuit test with $G=C Q$ noisy gates. U3 denotes general rotations on the Bloch sphere with angles $t(k)=\pm \pi/2, l(k)=\pm \pi/4, p(k)=\pm\pi/4$ and random signs. The CNOT gate entangles the state maximizing cascading effects.\label{randomalg1}}
\end{table}
\begin{figure}[ht]
\[ 
\Qcircuit @C=1em @R=1em {
 & \qw &  \gate{U_3} & \ctrl{1}& \qw        & \qw     &  \qw       & \qw      & \qw        & \targ & \qw \\ 
 & \qw &  \qw        & \targ   & \gate{U_3} & \ctrl{1}&  \qw       & \qw      & \qw        & \qw   & \qw\\
 & \qw &  \qw        & \qw     & \qw        & \targ   &  \gate{U_3}& \ctrl{1} & \qw        & \qw   & \qw\\
 & \qw &  \qw        & \qw     & \qw        & \qw     &  \qw       & \targ    & \gate{U_3} & \ctrl{-3} &\qw
}
\] 
\caption{Diagram of one of $C$ cycles of the random circuit test. Each cycle randomly rotates all qubits in the Bloch sphere with the rotation gate $U_3(\pm \pi/2, \pm \pi/4, \pm \pi/4)$ and random signs. It was found empirically that it produces a uniform distribution at $7$ cycles starting from $|0\rangle$.\label{cycle}}
\end{figure}
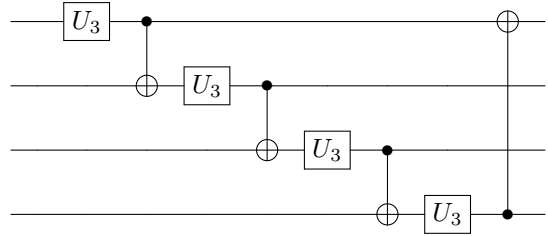

\begin{figure}[ht]
    \centering
    \includegraphics[width=9cm]{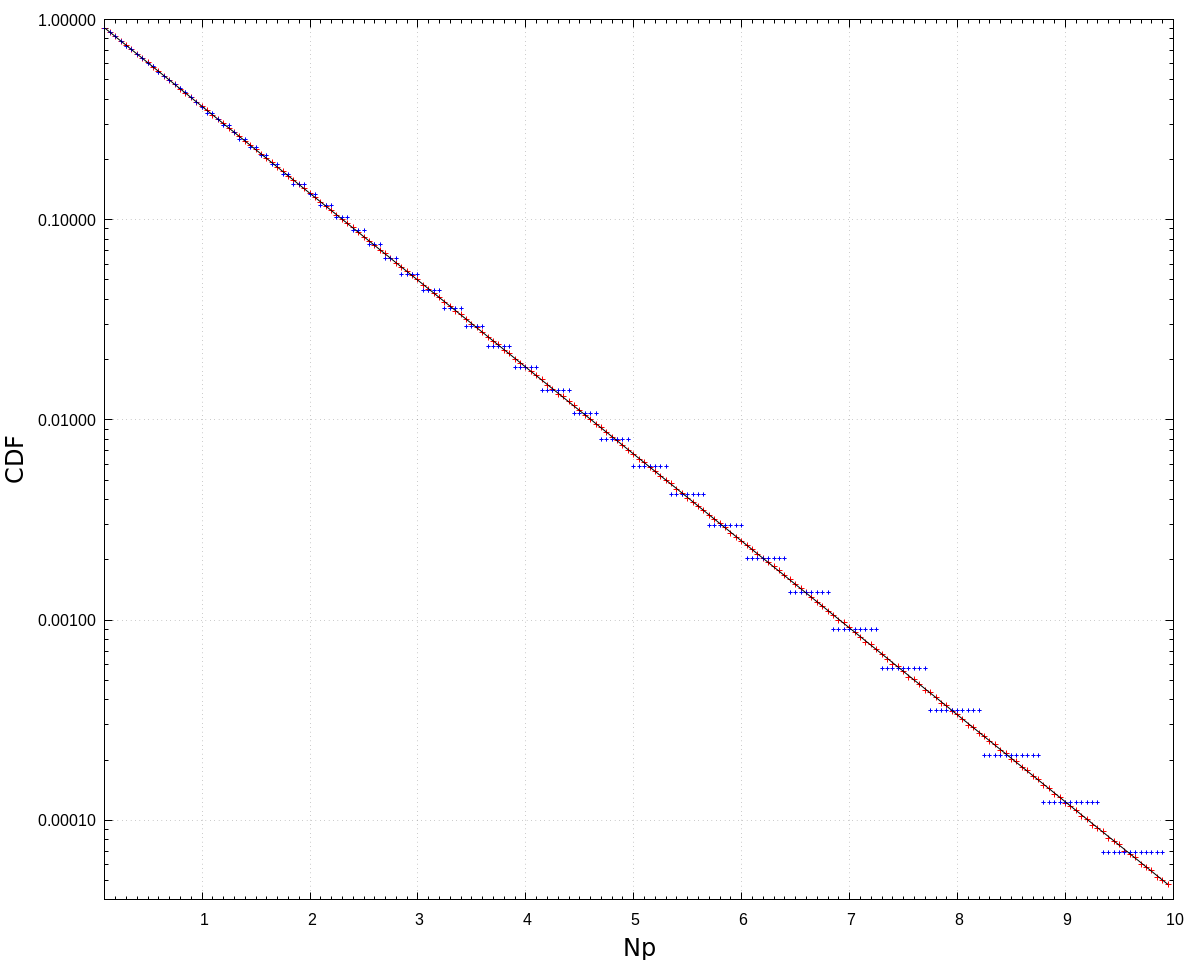}
    \caption{Cumulative probability function of $P(|c_k|^2>p)$ with the initial condition $|0\rangle$ after $7$ cycles and $Q=30$, 
        for $E=4, F=9, A=11$ (red crosses) and $E=4, F=5, A=7$ (stepped blue sums). The line is the cumulative exponential distribution $P(|c_k|^2>p)=e^{-Np}$.  
    \label{porterthomasfigure}      }
\end{figure}

The circuit is composed of $C$ cycles, each one randomly rotates all qubits in the Bloch sphere with the general rotation gate $U_3$,
\begin{equation}
U_3(\theta, \lambda, \phi)=\begin{pmatrix}
                                 \cos \frac{\theta}{2} & -e^{i\lambda}\sin \frac{\theta}{2}\\
                                 e^{i\phi} \sin \frac{\theta}{2}  &   e^{i(\lambda+\phi)}\cos \frac{\theta}{2}
\end{pmatrix} 	
\end{equation}
where the angles are $U_3(\pm \pi/2, \pm \pi/4, \pm \pi/4)$ with random signs at each gate application.
After each rotation a CNOT gate entangles the state. The results of each gate depend on the output of the gate before (Fig. (\ref{cycle})) to maximize cascading effects. It was found empirically that for $Q\le 30$ this circuit
generates a uniform distribution after approximately $7$ cycles starting from $|0\rangle$, the entropy is maximized and the resulting coefficients have statistically insignificant auto-correlations. Fig. (\ref{porterthomasfigure}) shows the cumulative distribution of probabilities $P(|c_k|^2>p)$
computed with double precision and low precision arithmetic, it converges to the exponential (Porter-Thomas) cumulative distribution 
\begin{equation}
P(|c_k|^2>p)= \int_p^\infty  N e^{-Ns} ds=  e^{-Np}.
\end{equation}
 The same graph 
shows the same run with low precision $E=4, F=9, A=11$ and for $E=4, F=5, A=7$, the truncation errors become evident as steps. 

To test the accumulation of errors we perform two experiments: a) start with a uniformly distributed random initial condition, and compare low precision versus double precision to test Eq. (\ref{iid}), and b) start with a non-random initial condition to show the scheme works for more general states. 

\begin{figure}[ht]
	\centering
	\includegraphics[width=8cm]{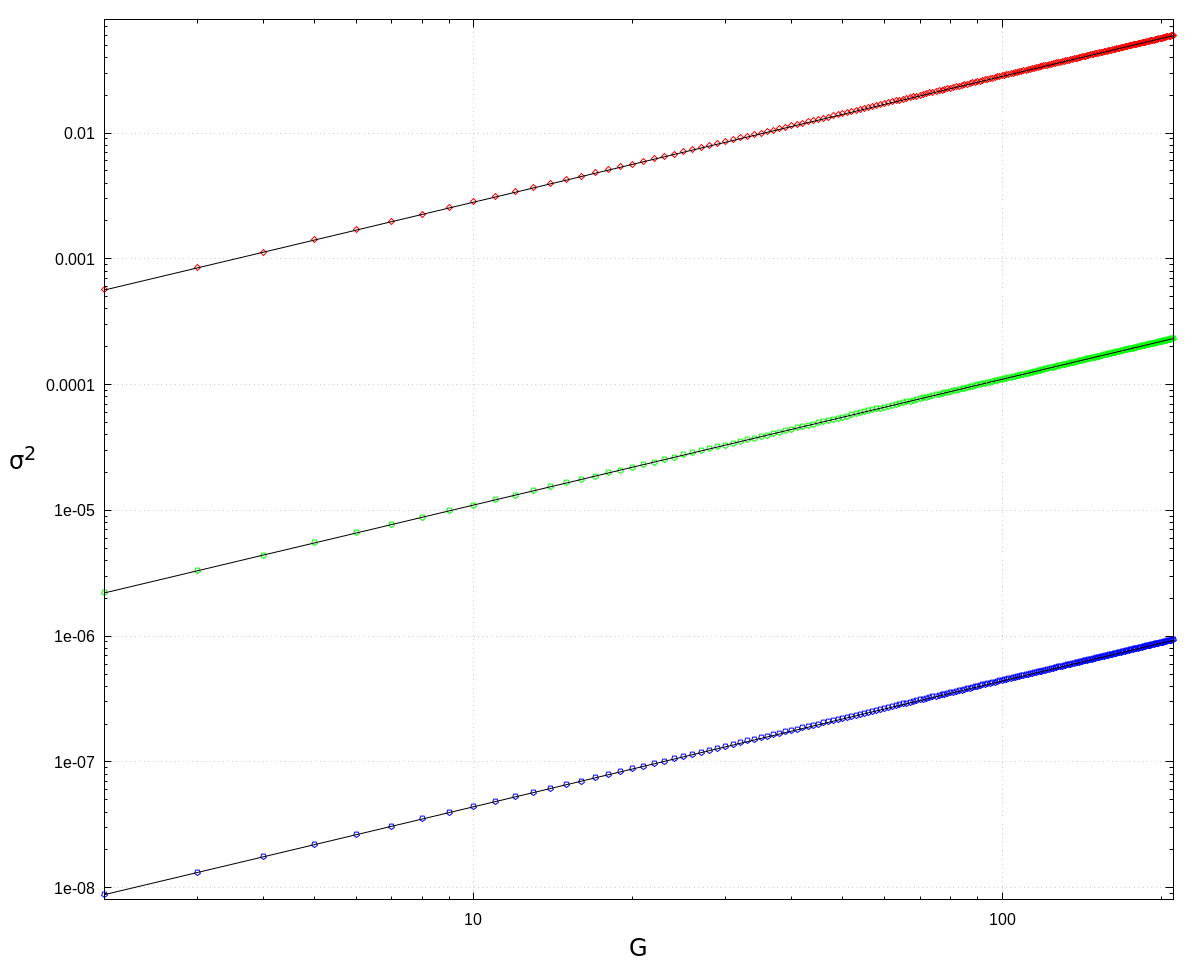} 
	\caption{Growth of the numerical cumulative error Eq. (\ref{cumulative}) (points) for a uniformly distributed, random initial condition, as a function of the number of error prone gates $G$, compared with Eq. (\ref{iid}) (lines), with $Q=30$ for triplets $E,F,G$: $4,5,7$ (top line),  $4,9,11$ (middle) and $4,13,15$ (bottom). The error is computed by comparing the output with low precision $|\psi_G \rangle $ with a computation with double precision as a proxy for the exact solution $|\psi_{ex} \rangle $. 
	\label{sigma-vs-G}}
\end{figure}

In the first experiment we start with a random initial condition 
\begin{equation}
c_k= \frac{\mathcal{N}(0,1)+i \mathcal{N}(0,1)}{K}
\end{equation}
where $\mathcal{N}(0,1)$ is a normal distribution with zero mean and unit variance, and $K$ is a normalization constant to enforce Eq. (\ref{normalization}). The real and imaginary components of this initial state are uniformly distributed on a sphere of $2N$ dimensions, and represents a maximally entangled state of high entropy.
Then we compute the amplitudes with $7$ cycles of the random algorithm (Table \ref{randomalg1}), first with double precision $|\psi_{ex} \rangle$ (as a proxy of the exact solution) and then with low precision arithmetic $|\psi_G \rangle $, and compute the error $\sigma^2=\|\; |\psi_G\rangle - |\psi_{ex}\rangle \; \|^2$. Fig. (\ref{sigma-vs-G}) shows agreement with the model for $Q=30$ and 3 sets of values of the triplets $E,F,A$.  

Fig. (\ref{histogramerrors}) shows the histograms of the cumulative errors of the real part of the coefficients, $\operatorname{Re}(c_{k, double}-c_{k, lowprec})$ for $E=4, F=9, A=11$ after $7$ cycles. The distribution is approximately normal with standard deviation $\sigma$ given by Eq. (\ref{iid}) (line). This adds support to the use of the central limit theorem in the derivation.
\begin{figure}[ht]
	\centering
	\includegraphics[width=9cm]{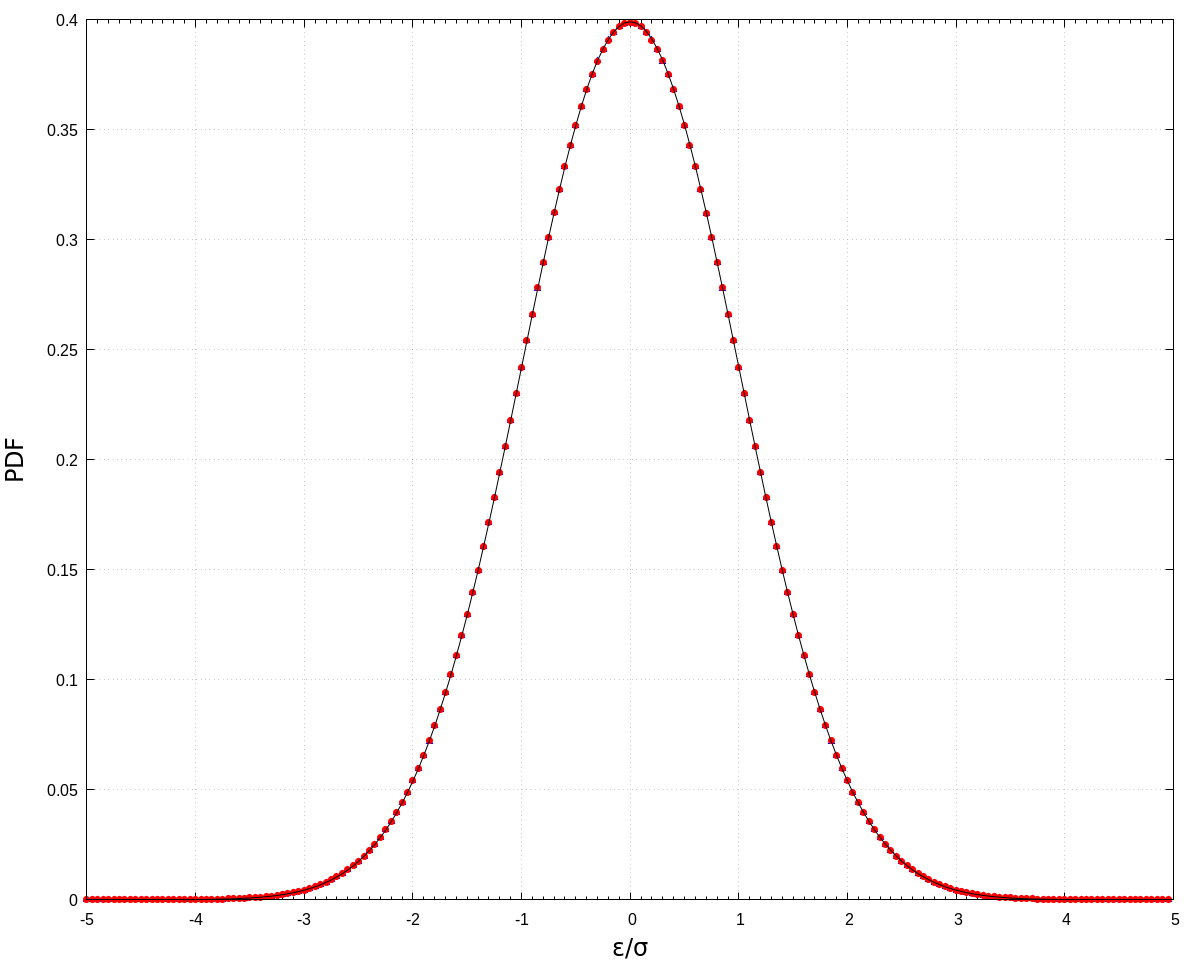}
	\label{histogramerrors}
	\caption{Starting with a uniform random initial condition we run $7$ cycles 
		twice, first with double precision and then with low precision. These are the histograms of the normalized errors of the real part of the coefficients, $\operatorname{Re}(c_{k, double}-c_{k, lowprec})$ for $E=4, F=9, A=11$ (points). The distribution is approximately normal with standard deviation $\sigma$ given by Eq. (\ref{iid}).
	\label{histogramerrors}}
\end{figure}

\begin{figure}[ht]
	\centering
	\includegraphics[width=8.5cm]{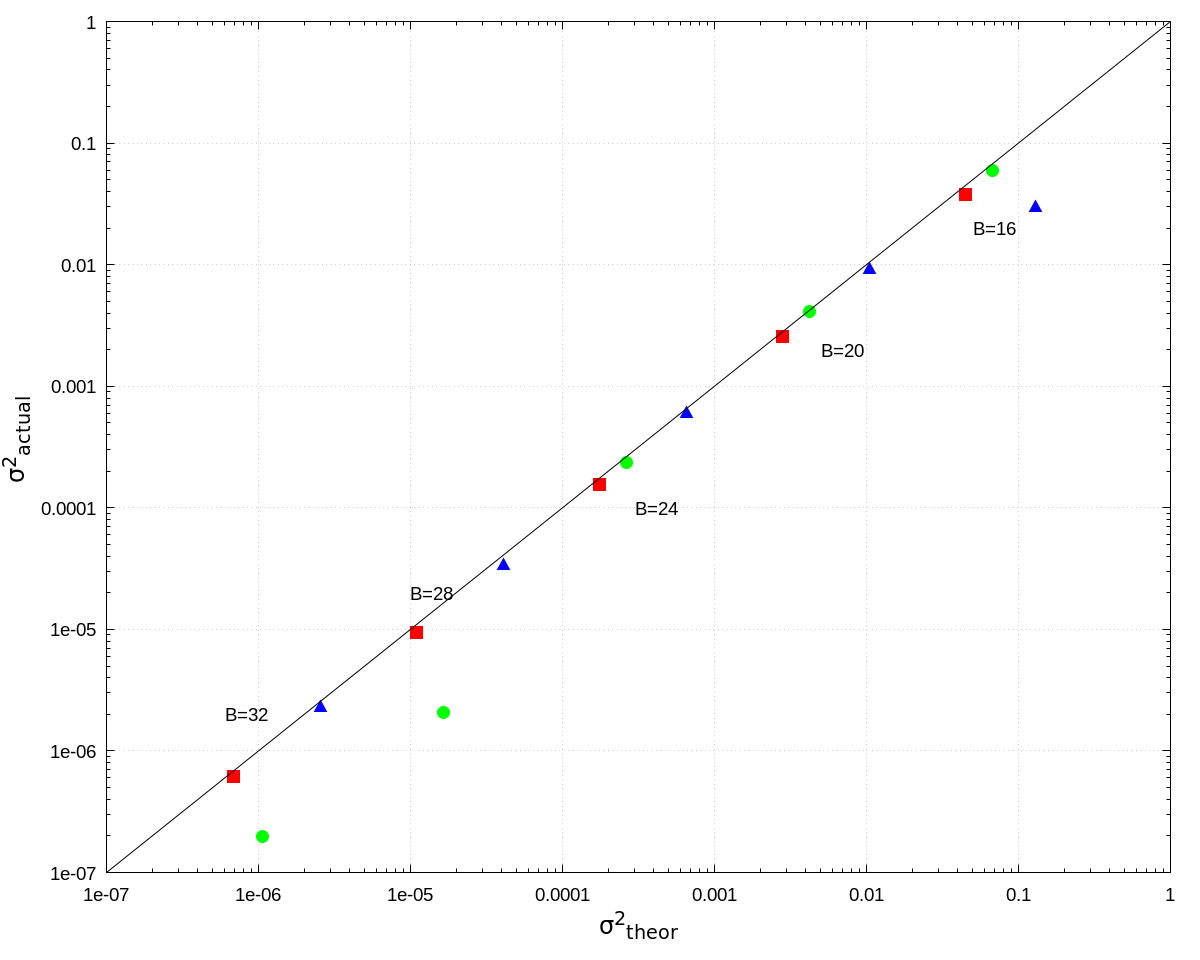} 
	\caption{Random algorithm test (Table \ref{randomalg2}), 4 cycles forth and then 4 cycles for the inverse. Comparison of the actual error (y-axis) and the theoretical error from Eq. (\ref{iid}) (x-axis). Red squares: 20 qubits, brown circles: 30 qubits, blue triangles: 40 qubits. The bits per coefficient are indicated on the labels, and the values of $E,F,A$ are from Table \ref{EFArandom}. \label{U3runs} 
	}
\end{figure}

In the second experiment, we test an initial condition as far as possible from a uniform distribution. We choose a random integer $x_0$ on $0\le x_0<N$ and set the initial condition 
\[ c_k= \begin{cases} 
1 & k=x_0 \\
0 & k\ne x_0 .
\end{cases}
\]
Then we run $C=4$ cycles of the random circuit, at which point the state is not yet a maximum entropy state, normalize with Eq. (\ref{normalize}), and finally run the inverse gates in reverse order, and compute the empirical error $\sigma_{actual}^2$ as
\begin{equation}
\sigma^2_{actual}= |c_{x_0} -1|^2+ \sum_{k=0,k\ne {x_0}}^{N-1} |c_k|^2.
\label{trueerror}
\end{equation}
(See Table \ref{randomalg2}). Fig. (\ref{U3runs}) compares the actual versus the theoretical error for various $Q$ and parameters $E,F,A$. 
 Even when during much of the computation the state is not uniformly distributed, the errors follow the model remarkably well.

We have also tried several variations of the random algorithm, such as using chained random gate controls with arbitrary topology, or arbitrary rotations on the interval $(0,\pi/2)$, and the results are the same.

\begin{table}[ht]
	\centering
	\begin{tabular}{|c|}
		\hline        
		\verb| // CREATE A RANDOM STATE                |\\                        
		\verb| for i=1,C                               |\\
		\verb|     for q=1,Q                           |\\
		\verb|         k= Q*i+q                        |\\
		\verb|         U3(q, t(k), l(k), p(k) )        |\\
		\verb|         CNOT(q, (q+1)%Q )               |\\
		\verb|     end                                 |\\
		\verb| end                                     |\\
		\verb| NORMALIZE                               |\\  
		\verb| // RUN IN REVERSE ORDER TO RESTORE IC   |\\
		\verb| for i=C,1                               |\\
		\verb|     for q=Q,1                           |\\
		\verb|         k= Q*i+q                        |\\
		\verb|         CNOT(q, (q+1)%Q )               |\\
		\verb|         U3(q, -t(k), -p(k), -l(k) )     |\\
		\verb|     end                                 |\\
		\verb| end                                     |\\  
		\verb| NORMALIZE                               |\\
		\hline
	\end{tabular}  
	\vspace{2mm}
	\caption{The non uniform initial condition test starts at state $|x_0\rangle$ and has $G=2 C Q$ noisy gates. U3 denotes general rotations on the Bloch sphere with angles $t(k)=\pm \pi/2, l(k)=\pm \pi/4, p(k)=\pm\pi/4$ with random signs. \label{randomalg2}}
\end{table}

\subsection{Quantum Fourier Transform}
The Quantum Fourier Transform \cite{Shor} (QFT) is an interesting algorithm because it contains many gates unresolved by the low precision format. QFT consists on a series of Hadamard gates applied to each qubit $p=1,2..Q$, each one followed by controlled phases to all qubits $q>p$ with phase $\pi/2^{q-p}$, and finally the order of the qubits is inverted with swaps at the end. Thus if $q-p \ge A$ the phase gates are unresolved and the error-canceling mechanism from Eq. (\ref{argcorrection}) is necessary. 

We verified that by feeding uniformly distributed random states to the QFT then the results are in good agreement with the model. This is not surprising because this initial data satisfies all the conditions for the derivation of the errors in section III, and also it is analogous to the first experiment we did with random circuits. From the point of view of comparing the errors, a more interesting test is to feed non-random data to the QFT (although this is an easy test for the tensor contraction formulation). 

\begin{figure}[ht]
    \centering
    \includegraphics[width=8.5cm]{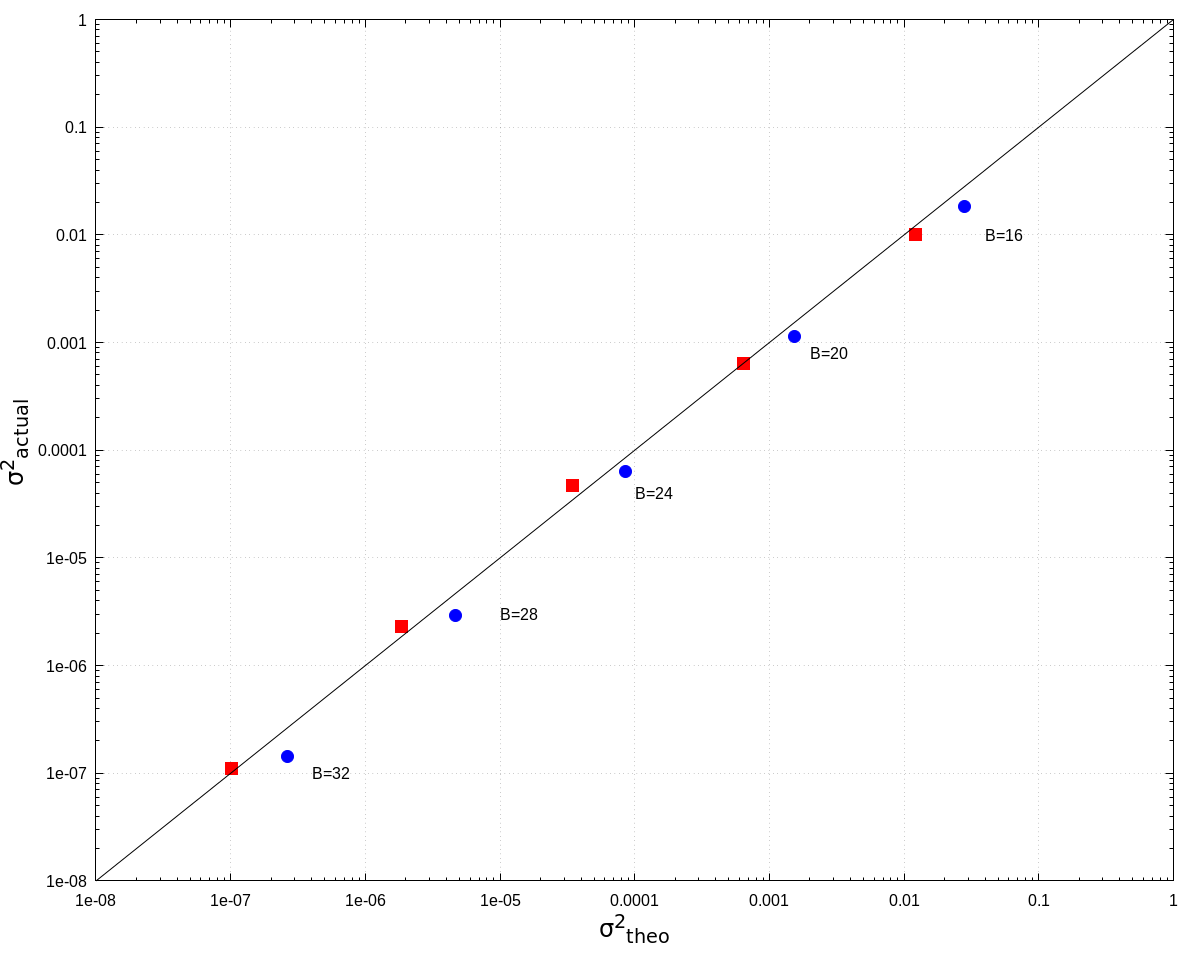} 
    \caption{Quantum Fourier transform test: cumulative error as a function of theoretical value Eq. (\ref{iid}) for $Q=20$ (red squares) and $Q=30$ (blue circles), and triplets $(E,F,A)$: $(4,5,7)$, $(4,7,8)$, $(4,9,11)$, $(4,11,13)$ and $(4,13,15)$. 
    \label{QFT}}
\end{figure}

For this test, first we choose a random integer $0\le x_0<N$ and the initial amplitudes are set equal to 
\begin{equation}
c_k= \frac{1}{\sqrt{N}}\exp\left( -\frac{2\pi i k x_0}{N} \right),
\end{equation}
then the QFT is computed with one normalization at the beginning and one at the end, and then the true errors $\sigma^2_{actual}$ are computed as before with Eq. (\ref{trueerror}).
Fig. (\ref{QFT}) compares the true errors with the theoretical errors Eq. (\ref{iid}) for $Q=20$, $Q=30$ and various values of $E,F,A$. Notice that even in the data used in this test is not random, the agreement with the model is reasonable.

For some applications such as factorization of integers, it has been shown that the Approximate Quantum Fourier Transform (AQFT) requires fewer phase gates \cite{Coppersmith,Barenco} than the QFT. In this approximation, instead of applying the phase gates to all qubits $q>p$,  only $m=\lfloor 3+\log_2 Q \rfloor$ phases are sufficient with $q-p\le m$. For example for $Q=50$, $m$ is 9, so if the algorithm runs on low precision arithmetic with $A\ge 10$ (23 bits/amplitude according to Table \ref{EFArandom}) then the phase gates do not contribute any error and only the Hadamard gates do.

\section{Summary and open questions}

We developed a model for the accumulation of errors on quantum circuit simulations with a low precision logarithmic representation for the amplitudes and a full vector state formulation. Each coefficient uses $B=E+F+A$ bits, where $E$ and $F$ are the bits of the integer and fractional parts of the logarithm and $A$ the bits of the complex argument.

\subsection{Main findings}
\begin{itemize}

\item For maximally entangled random states, the cumulative quadratic error for a circuit with $G$ error prone gates can be approximated as
\[
\sigma^2 \approx \varepsilon^2_c G
\] 
where $\varepsilon^2_c$ is the total conversion error,
\[
\varepsilon^2_c  \approx \phi+(1-\phi)\frac{2^{-2F}+4\pi^2 2^{-2A}}{12},
\]
$\mu=\exp(-2^E+2^{-F})$ is the smallest representable modulus and $\phi=1-(N\mu^2+1)e^{-N\mu^2}$ is the loss of normalization per conversion due to underflow.

\item This expression can be used to estimate the maximum number of error prone gates that can be run before the error grows above $\sigma^2$ for random states,
\[
G_{random}< \frac{\sigma^2}{ \varepsilon^2_c}.  
\]
Table \ref{Gtable} shows typical values predicted for different values of $B$, values as low as $B=16$ can yield usable results for hundreds of gates with 50 qubits. This is significant because word sizes $16 \le B \le 32$ represent memory and bandwidth saving factors of approximately $4X$ to $8X$ with respect to double precision arithmetic (128 bits per amplitude).

\item Not all gates are susceptible to rounding errors, and the effective number of error prone gates can be approximated by $G=\sum_{g=1}^{G_0} \beta_g$ with the weights $\beta_g\le 1$ given in Table \ref{effectiveG}. This number can be significantly smaller than the total number of gates $G_0$.

\item The optimal choices for the triplets $E,F,A$ are found by minimizing the error $\varepsilon_c^2$, and are summarized in Table \ref{EFArandom}. The angular number of bits is related to the fraction bits as $A= F+2$ or $A=F+3$ depending on the value of $B$. The table shows that $E=5$ is sufficient for the foreseeable future.

\item For non-random states and biased errors we can estimate an upper bound for the error, which grows quadratically with $G$,
\[
\sigma^2   \lesssim G^2\left( N\mu^2 + \frac{2^{-2F}+4\pi^2 2^{-2A}}{4} \right)
\]
The optimal choices for the triplets $E,F,A$ are found by minimizing the errors, and are summarized in Table \ref{EFAbiased}.

\item The biased case can be partially remediated by multiplying the coefficients by carefully crafted random factors and shown in sections \ref{biased1} and \ref{biased2}.
According to experiments performed by the author, provided the normalization procedure is used, the actual errors are closer to the random case than to the upper bound.

\item The examples with random circuits and the QFT show that the model based on random states Eq. (\ref{iid}) may still be a good approximation even if the amplitudes are not random.

\end{itemize}

\subsection{Open questions}
\begin{itemize}

\item An important theoretical question is whether this format is optimal for the simulation of maximally entangled random states with Schrodinger's formulation, that is, whether there is another format that uses less storage $B$ for a given error $\sigma^2$. 

\item Another issue is optimization: to make the low precision format practical, the conversion between low precision and double precision needs to be optimized for speed, either with software or with hardware (if the simulation is done with FPGA's, for example). 

\end{itemize}

\section{Acknowledgments}
Many thanks to Datavortex Technologies that supported this work and provided the computer system Hypatia used in most of the development, to the Texas Advanced Computing Center (TACC), University of Texas at Austin, for providing access to the Stampede 2 system, and the University of North Texas HPC Research IT Services for providing access to TALON 3 computer.

\end{document}